\documentclass[review]{elsarticle}

\usepackage{hyperref}
\usepackage{graphicx}

\usepackage{filecontents}
\usepackage{amsmath,amssymb,amsfonts,tabu}
\usepackage{caption}
\usepackage{textcomp}
\usepackage{xcolor}
\def\BibTeX{{\rm B\kern-.05em{\sc i\kern-.025em b}\kern-.08em
    T\kern-.1667em\lower.7ex\hbox{E}\kern-.125emX}}

\journal{Big Data Research}

\begin{document}

\begin{frontmatter}

\title{Online Subspace Tracking for Damage Propagation Modeling and Predictive Analytics :Big Data Perspective}

\author{Farhan Khan}
\address{National University of Sciences and Technology, Islamabad}
\ead{farhan.khan@seecs.edu.pk}

\begin{abstract}
We analyze damage propagation modeling of turbo-engines in a data-driven approach. We investigate subspace tracking assuming a low dimensional manifold structure and a static behavior during the healthy state of the machines. Our damage propagation model is based on the deviation of the data from the static behavior and uses the notion of health index as a measure of the condition. Hence, we incorporate condition-based maintenance and estimate the remaining useful life based on the current and previous health indexes. This paper proposes an algorithm that adapts well to the dynamics of the data and underlying system, and reduces the computational complexity by utilizing the low dimensional manifold structure of the data. A significant performance improvement is demonstrated over existing methods by using the proposed algorithm on CMAPSS Turbo-engine datasets.
\end{abstract}


\end{frontmatter}

%\linenumbers

\section{Introduction}
\label{intro}
Machine prognostics and predictive analytics are widely investigated in control theory, industry applications and signal processing \cite{abbas_hierarchical_2009,wang_similarity-based_2008,javed_new_2015,eker2014similarity}. Condition based maintenance uses prognostics methods in a variety of applications such as manufacturing, automotive, heavy industry, consumer electronics and biomedical equipments. The correct estimate of future condition is helpful and add to the timely maintenance/replacement of the faulty component(s). In such manner predictive analytics helps save time, effort and cost and assure the smooth running of the required processes. In literature, various methods have been studied using physical modeling of the degradation. However, physical modeling in most modern applications are inadequate and at times extremely complex and stochastic. Therefore, more recently data-driven methods are investigated that use instantaneous sensors' as well as operational data collected from the machines \cite{ramasso_performance_2014}. To this end, we propose a novel data-driven algorithm that analyze time-series data for the degradation modeling and remaining useful life (RUL) estimation.

Various data-driven methods have been investigated in literature for damage propagation modeling and predictive analysis specifically in machine learning, signal processing, time-series analysis and deep learning \cite{babu_deep_2016,heimes_recurrent_2008}. Conventionality, the sensors' data from machines through their life cycle (till failure) is used to train the damage propagation model and estimate remaining useful life of new instances. In such scenario, the RUL is represented by a linear decaying function, i.e., $RUL_t=T-t$, where $T$ is time at failure. However it's more practical to assume a piece-wise linear function since, in the beginning when machine is operating in perfectly healthy condition, the health of machine cannot be taken as degrading. For that purpose, the RUL is approximated as \cite{zheng_long_2017}:
\begin{equation}
\begin{split}
RUL_t&=T ~~~~~~~~for~ 0 \leq t \leq t_d \\
	 &=T-t~~~~~for~ t_d < t \leq T,
\end{split}
\end{equation}
where $t_d$ is the point where the linear degradation starts. The piece-wise RUL is then use as target in supervised learning during the training phase of the algorithm. Several algorithms such as support vector regression (SVR), Convolutional neural networks (CNN) and more advanced deep learning involving Long Short-Term Memory (LSTM) and CNN have been used for degradation modeling and RUL estimation \cite{zheng_long_2017}. However, here we investigate a semi-supervised approach where no assumption is made with regards to RUL and the degradation modeling is learned from the input data entirely as in \cite{wang_similarity-based_2008,eker2014similarity,malhotra_multi-sensor_2016,wang_trajectory_2010}. To this end, we assume that the time till the machine runs in healthy state is known that can be incorporated in a novelty detection model \cite{malhotra_multi-sensor_2016}. In this manner, the deviation in statistical distribution of the data is used as a measure of the degradation. 

For damage propagation and performance degradation modeling, various methods are investigated in literature that mainly involve auto-encoders for input sequence reconstruction \cite{malhotra_multi-sensor_2016}. These methods rely on the reconstruction error as a measure of the damage (and hence) health of the machine at any given time. For instance, in \cite{malhotra_multi-sensor_2016}, the authors use LSTM based encoder-decoder model that regenerates the input sequence, after training on the healthy samples, and use the error between estimated and true input sequence as measure of the degradation and health index. Furthermore, they employ linear regression along-with the encoder-decoder model for more robust modeling. However, we propose a subspace tracking approach to measure the variation in the distribution of input sequence by incorporating instantaneous manifold tracking \cite{lin_riemannian_2008,khan_universal_2016}. In the proposed approach, instead of regenerating the whole input sequence, we estimate a low dimensional representation of the input and the subspace that it lies in, hence reducing the computational cost and overfitting. Furthermore, since the input and underlying submanifold subspace have different dimension, we incorporate approximate Mahalanobis distance for updating the model parameter during training and later as a measure of degradation \cite{xie_online_2013,xie_change-point_2013,wang_adaptive_2005,khan_universal_2016}. We emphasize that our proposes algorithm suits well to the dynamics of the input data, reduces the computational complexity and achieves significantly higher accuracy than the state-of-the-art. We demonstrate the performance of the proposed algorithm by applying it to the well known CMAPSS datasets \cite{liu_users_2012,ramasso_performance_2014}.    

In summary, we investigate online subspace tracking for the damage propagation assuming low dimensional manifolds. We further incorporate linear regression by using the low dimensional projection of the input data as new input and estimate the health index. We then estimate the RUL of a test case by comparing the health index curve to all available degradation models. We use an ensemble learning approach as in \cite{malhotra_multi-sensor_2016} to finally estimate the RUL. The organization of the paper is as follows: in Section \ref{sec:2}, we formally describe problem setting and define various parameters in detail. In Section \ref{algo}, we demonstrate the subspace tracking algorithm for damage propagation modeling and health index curves generation. In Section \ref{results}, we apply the proposed damage propagation modeling to real life datasets and demonstrate the performance evaluation in terms of root mean square error (RMSE) and the scores defined in \cite{ramasso_performance_2014}. We finally conclude the paper in Section \ref{conclusion}.
\section{Problem Description}\label{sec:2}
All vectors used in this paper are column vectors denoted by boldface lowercase letters. Matrices are denoted by  
boldface uppercase letters. For a vector  $ \mathbf{x} $ (or a matrix $ \mathbf{U} $), $ \mathbf{x}^T $ ($ \mathbf{U}^T $) 
is the ordinary transpose. $ T $ is the total number of time-steps and an arbitrary 
time-step is denoted by $ t $ where $ 0\leq t < T-1$. The time index of a sequence vector is denoted by $ t $ in the subscript, as in $\mathbf{x}_{t}$.

We investigate the estimation of current health index (HI) $\sigma_t$ and the prediction of remaining useful life $RUL_{t}$ by analyzing the input data $\mathbf{x}_t \in {\rm I\!R}^D$ where $D$ is the number of input features. The health index is modeled as a function of the input data as:
\begin{equation}
\sigma_t = f(\mathbf{x}_t),
\end{equation}
where $0\leq\sigma\leq1$, with $1$ and $0$ corresponding to perfect health and failure respectively. Furthermore, we assume that the $D-$dimensional input data lies on a static or time varying submanifold, $S_{k,t}$ with reduced intrinsic dimension $d$ such as $d \ll D$, where, for the multiscale modeling of a non-stationary submanifold, $k$ is the index such as $k \in \{1,2,...,K\}$ \cite{xie_change-point_2013,xie_online_2013}. 
\begin{figure}[ht!]
	\centering
	\includegraphics[width=\linewidth,height=\textwidth]{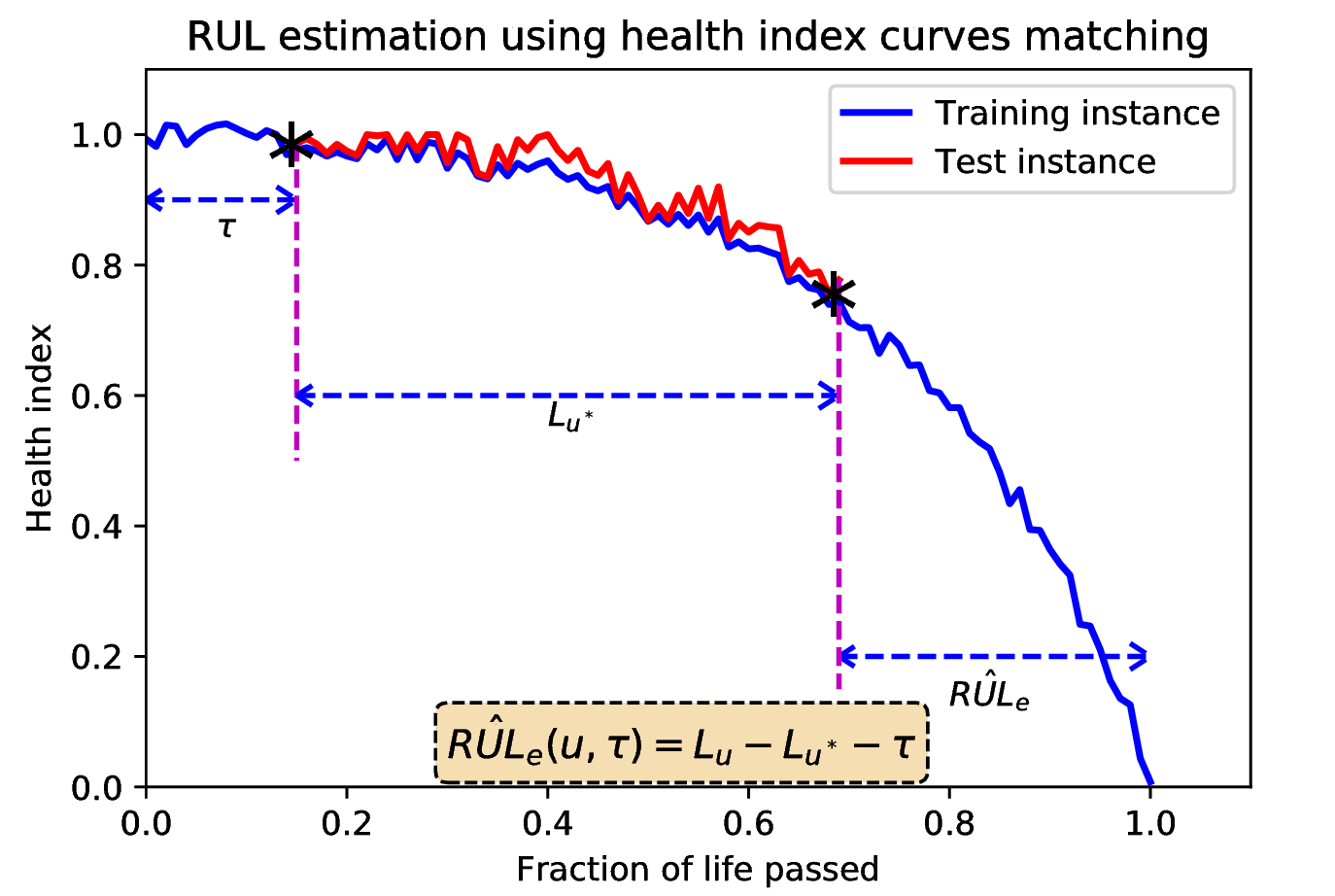}
	\caption{Health index curve and RUL estimation using similarity}
	\label{fig:rulhi}
\end{figure}

\subsection{Health index and Remaining Useful Life Estimation}\label{hi-rul}
Given HI curves for the all training instance, we use curve matching for the estimation of RUL \cite{eker2014similarity,wang_similarity-based_2008,wang_trajectory_2010}. That is, as shown in Fig. \ref{fig:rulhi}, the HI trajectory for a test instance $u^*$ is slided over training HI curve sets $U$ for all times $\tau_1 \leq \tau \leq \tau_2$ in order to minimize the euclidean distance. The similarity between test HI curve $u^*$ and training HI curve $u$ is defined by \cite{wang_similarity-based_2008,eker2014similarity,wang_trajectory_2010}:
\begin{equation}
	s(u^*,u,\tau) = \exp(-d^2(u*,u,\tau)/\beta),
\end{equation}
where $\tau \in \{\tau_1,...,\tau_2\}$ is the time-lag, $\beta > 0$ is a small constant that controls the similarity and $d^2(u^*,u,\tau)$ is the squared euclidean distance defined by,
\begin{equation}
d^2(u^*,u,\tau) = \frac{1}{L_{u^*}}\sum_{i=1}^{L_{u^*}}(\sigma^{u^*}_i - \sigma^{u}_{i+\tau})^2.
\end{equation}
For each training instance $u$ and time-lag $\tau$, the RUL of test instance $u^*$ is estimated as: $RUL_{u^*}(u,\tau) = L_u - L_{u^*} - \tau$. The final estimate of RUL is a linear combination of $RUL_{u^*}(u,\tau)$ all $u,\tau$ using the similarity measure $s(u^*,u,\tau)$ as coefficients, i.e.,
\begin{equation}
\hat{RUL}_{u^*} = \frac{\sum s(u^*,u,\tau) \sum RUL_{u^*}(u,\tau)}{\sum s(u^*,u,\tau)}. \label{rul}
\end{equation}

The error $e_{u^*} =\hat{RUL}_{u^*} - RUL_{u^*}$ between estimated and true RUL on test sets is used as a measure of performance. We specifically use two performance metrics, i.e., the RMSE defined as,
\begin{equation}
RMSE = \sqrt{ \sum_{u^*=1}^{N}(\hat{RUL}_{u^*} - RUL_{u^*})^2},
\end{equation}
and score $S$ defined as,
\begin{equation}
S = \sum_{u^*=1}^N (\exp(\frac{\gamma}{e_{u^*}})-1),
\end{equation}
where $\gamma = 1/13$ when $e_{u^*}<0$ and $\gamma = 1/10$ when $e_{u^*}\geq 0$. This way a late detection of the failure (smaller estimated RUL) is penalized more as compared to early detection \cite{ramasso_performance_2014}.

\section{Multi-scale Subspace Tracking for Predictive Analytics} \label{algo}

In the most basic form, we propose to use a single subspace tracking algorithm for the damage propagation modeling. In this sense, the input data $\mathbf{x}_t \in {\rm I\!R}^D$ in healthy state is assumed to be lying on a static submanifold with intrinsic dimension $d$. We project the input data on the subspace $S_t$ of the submanifold and then determine the approximate Mahalanobis distance $d(x,S)$ as in \cite{xie_change-point_2013}. The square root of distance is used as error to update the parameters of estimated subspace $\{\mathbf{U},\mathbf{c},\mathbf{\Lambda}\}$. Here matrix $\mathbf{U}$ is the eigen-vector matrix of the covariance matrix representing the orientation of the subspace, vector $\mathbf{c}$ is the mean of input space and $\mathbf{\Lambda}=diag\{\lambda_1,...,\lambda_d\}$ represents the spread. 
	
The subspace tracking is run on the healthy data in epochs till the error is minimized and converged while using the instantaneous error to update the parameters in a stochastic gradient descent manner. Once the subspace parameters are finalized, the trained model is used on the remaining cycles of data till failure and the approximate Mahalanobis distance is recorded for each time instant as \cite{xie_online_2013,khan_universal_2016}:
\begin{equation}\small{
	d_t (\mathbf{x},S) \triangleq \delta (\mathbf{x} -\mathbf{c})^T \mathbf{U}_1 \mathbf{\Lambda}_1^{-1} \mathbf{U}_1^T (\mathbf{x} -\mathbf{c}) + 
	\rVert \mathbf{U}_2^T (\mathbf{x}-\mathbf{c})\rVert^2,} \label{Mahalanobis}
\end{equation}
where $\delta>0$ is a small constant that depends on the distribution of data beyond the submanifold. 
We then use this distance to generate health index $\sigma_t$ as:
\begin{equation}
\sigma_t = 1-\sqrt{\hat{d}_t(\mathbf{x},S)},
\end{equation}
where $0 \leq \hat{d}_t(x,S)\leq 1$ is the scaled version of $d(x,S)$. We get an exponentially decaying curve of the health index values that reaches $0$ at failure as shown in Fig. \ref{fig:rulhi}. 

To this end, we use subspace tracking of the $d-$dimensional submanifold and use the tracking error as measure of degradation for all training instances. However, after using a certain amount of data for generating HI curves, we next utilize linear regression for the remaining instances and cycles to estimate the health index. In other words, we use the known HI values (estimated through subspace tracking) as target and learn the linear regression model by least squares method, then estimate the HI values for the remaining data. Here, instead of using the original $D-$dimensional input, we use the $d-$dimensional projections of the data on the submanifold subspace as the new input. In this manner, we further reduce the computational cost of the overall algorithm. 

For non-stationary setting, i.e., when the input data does not follow a static distribution in the healthy state, we assume the data lies on a time varying submanifold and use a multi-model learning of the underlying subspace. We use the notion of multi-scale tracking and MOUSSE algorithm as in \cite{xie_change-point_2013,khan_universal_2016,xie_online_2013}. In multi-scale subspace tracking, the input space is partitioned into $K$ regions where there is a different subspace representing each individual submanifold. The input data $\mathbf{x_t}$ at time $t$ is projected on each subspace and the one with minimum distance is used and updated for the next cycle. Also, for the health index calculation, the minimum distance is used as a measure of degradation. 
\begin{figure*}[ht!]
	\centering
	\includegraphics[width=\linewidth,height=0.85\linewidth]{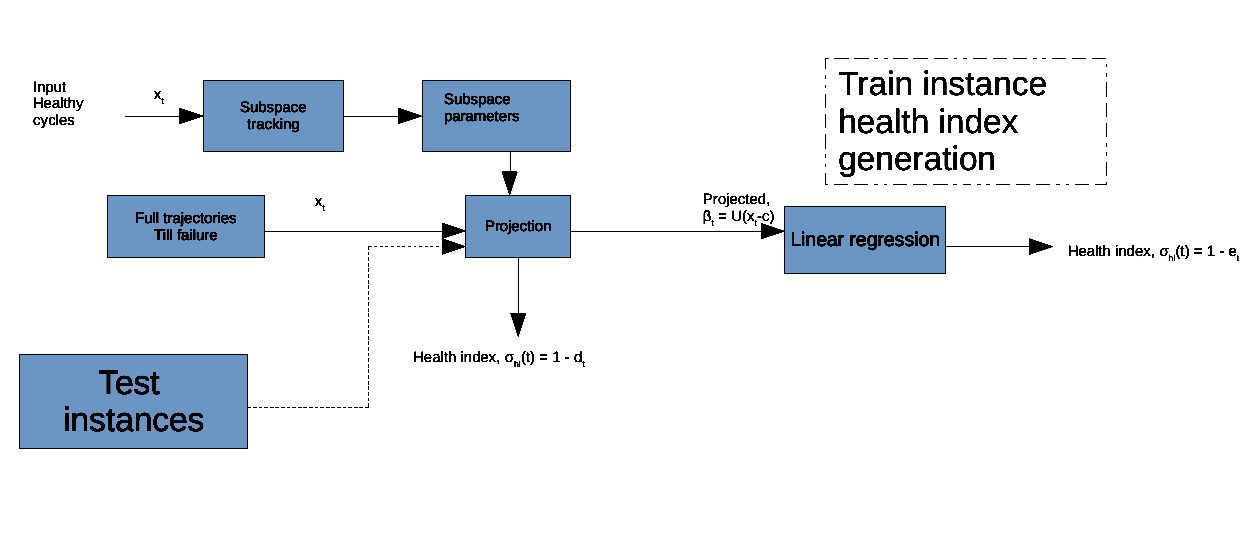}
	\caption{The proposed learning model for health index generation}
	\label{fig:model}
\end{figure*}\\
\begin{figure}[t!]
	\centering
	\includegraphics[width=0.9\linewidth,height=0.9\textwidth]{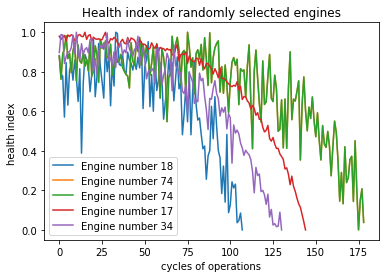}
	\caption{Health index curves of FD001 training dataset}
	\label{fig:figure1}
\end{figure}

\subsection{Algorithm Description}\label{AA}
Here, we briefly describe the algorithm shown in Fig. \ref{fig:model} step by step. Initially, we use data from the healthy state and train the subspace tracking algorithm till convergence, i.e., when the health index, $\sigma_t=1$. Then during the second stage (inference), we use the estimated parameters of the underlying subspace to estimate health index for the remaining cycles till the failure point. In this manner, we get an exponentially decaying curve. Furthermore, we use the trained model to estimate $\sigma_t$ for the test instances that are truncated before the failure. Finally, we use similarity measures for curve matching for the estimation of RUL as described in subsection \ref{hi-rul}.

\section{Results and Performance Analysis}\label{results}
\begin{figure}[th!]
	\centering
	\includegraphics[width=0.9\linewidth,height=0.65\textwidth]{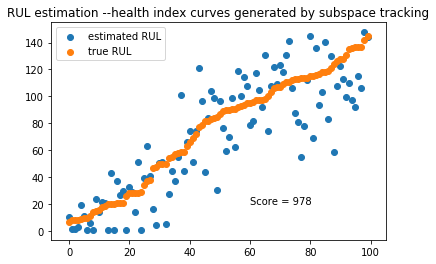}
	\caption{Subspace tracking (SST) without linear regression}
	\label{fig:figure9}
	\includegraphics[width=0.9\linewidth,height=0.65\textwidth]{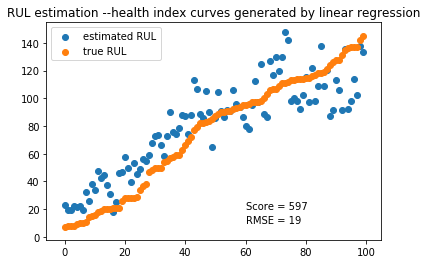}
	\caption{SST with linear regression (SST-LR)}
	
	\label{fig:figure93}
\end{figure}
To validate the proposed algorithm, we use the widely investigated CMAPSS turboengine datasets as benchmark \cite{liu_users_2012,ramasso_performance_2014}. In principal, the CMAPSS datasets consist of four independent datasets for various number of engines and time instances, where each one is a pair of training (for a complete cycle till failure) and test (where the data is truncated at a point before the failure point). The target is to estimate the remaining useful life of the test instances based on the behavior of the training data. The four datasets (in pair) are named as FD001, FD002, FD003 and FD004 in literature and each input consists of $24$ features that include $3$ operational setting features and $21$ sensor values. In all of the experiments, we assume the first $20$ cycles of each engine as healthy and employ subspace tracking. We use grid search cross-validation to choose the hyper-parameters as $\alpha=0.87, \tau_1=1, \tau_2=40,$ and $\beta=0.0235$. The intrinsic dimension of submanifold is set as $d=3$. We use three variations of the algorithm, i.e., Single subspace tracking (SST) for all HI curves generation, Single subspace tracking with linear regression (SST-LR) and multiple subspace tracking where the input data in non-stationary during the healthy state.

As second stage of the algorithm (inference), we plot the health index curves for all training and test instances. Fig. \ref{fig:figure1} shows health index curves for randomly selected five engines where $\sigma_t$ reaches zero at the end-of-life. Similar to \cite{heimes_recurrent_2008}, we match the HI curve for a certain test instance with all training instance curves and use \eqref{rul} to estimate RUL. In Fig. \ref{fig:figure9} and Fig. \ref{fig:figure93}, the estimated RUL for each test instance of dataset FD001 are plotted in the ascending order, using HI curves generated by SST and SST-LR respectively. The results show a good match between the true and estimated RUL, specifically with SST-LR. 

\begin{table}[t]
\caption{Turbofan Engine: Performance comparison w.r.t scores} \label{score-tab}
		\centering
	\resizebox{0.8\textwidth}{!}{$\begin{tabu} {|c|c|c|c|c|} \hline
		
		Algorithm & FD001 & FD002 & FD003 & FD004 \\ \hline \hline
		SVR & 1380 & 5.90\times10^5 & 1603 & 3.71\times10^5  \\ \hline
		CNN& 1290 &1.36\times10^4 & 1602& 7892 \\ \hline
		Deep LSTM &338 & 4452 & 852 & 5554\\ \hline
		LSTM-ED & 1260 & -- & -- & -- \\ \hline
		\textbf{SST} & 978 & \mathbf{4230} & \mathbf{822} & \mathbf{4401}\\ \hline
		\textbf{SST-LR} & 597 & \mathbf{3351} & \mathbf{634} & \mathbf{3381}\\ \hline
	\end{tabu}$}
\vspace{2em}	
\caption{Turbofan Engine: Performance comparison w.r.t RMSE values} \label{rmse-tab}
	\centering
	\resizebox{0.8\textwidth}{!}{$\begin{tabu}
	{|c|c|c|c|c|} \hline
		
		Algorithm & FD001 & FD002 & FD003 & FD004 
		\\ \hline\hline
		SVR & 20.96 & 42.00& 21.05 & 45.35  \\ \hline
		CNN& 18.45 &30.29& 19.82 & 29.16\\ \hline
		Deep LSTM &16.14 & 24.49 & 16.18 & 28.17\\ \hline
		LSTM-ED & 23.36 & -- & -- & -- \\ \hline
		\textbf{SST} & \mathbf{16.22} & 30.21 & 17.02 & 28.21\\ \hline
		\textbf{SST-LR} & \mathbf{15.02} & 29.12 & 16.95 & \mathbf{26.03}\\ \hline
	\end{tabu}$}
	
\end{table}

We apply the proposed algorithms on the remaining three datasets while using the multiple subspace tracking as described in subsection \ref{algo}. Specifically, by clustering the operational setting data, we observe that there are six different scenarios in case of datasets FD002 and FD004. This makes these datasets ideal candidate for multiple subspace tracking as in each operational case, the data lies on a one of the six submanifolds. Similarly, while using linear regression for HI values, we train a different model for each case. We compare the performance of proposed algorithms with that of support vector regression (SVR), Convolutional neural networks (CNN), LSTM based deep learning (Deep LSTM) \cite{babu_deep_2016,zheng_long_2017} and LSTM based encoder-decoder model (LSTM-ED) \cite{malhotra_multi-sensor_2016} as shown in Table I and Table II. The use of multi-model analysis makes the algorithm more robust and achieves significant performance improvement over SVR and CNN, and competes well against Deep LSTM while using a reduced computational complexity.

\section{Conclusion} \label{conclusion}

We investigate online manifold learning and subspace tracking for the damage propagation modeling. We propose a novel algorithm that generates health index values for each input based on the distribution of the data. We then use the HI curves for the estimation of RUL. We specifically investigate a damage propagation model of Turbo-engine, however, the proposed algorithm can be extensively applied to other applications for predictive analysis. The proposed algorithm is computationally efficient and adapts to the dynamics of the data both in static and non-stationary scenarios. We implement the proposed algorithm for the RUL estimation of CMAPSS Turbo-engines and achieves a significant performance improvement over the state-of-the-art in terms of RMSE and timely detection of the damage. 

\section*{References}

\bibliography{bigdata}

\begin{thebibliography}{10}
\expandafter\ifx\csname url\endcsname\relax
  \def\url#1{\texttt{#1}}\fi
\expandafter\ifx\csname urlprefix\endcsname\relax\def\urlprefix{URL }\fi
\expandafter\ifx\csname href\endcsname\relax
  \def\href#1#2{#2} \def\path#1{#1}\fi

\bibitem{abbas_hierarchical_2009}
M.~Abbas, G.~J. Vachtsevanos, A hierarchical framework for fault propagation
  analysis in complex systems, in: {AUTOTESTCON}, 2009 {IEEE}, {IEEE}, 2009,
  pp. 353--358.

\bibitem{wang_similarity-based_2008}
T.~Wang, J.~Yu, D.~Siegel, J.~Lee, A similarity-based prognostics approach for
  remaining useful life estimation of engineered systems, in: Prognostics and
  Health Management, 2008. {PHM}. International Conference on, {IEEE}, 2008,
  pp. 1--6.

\bibitem{javed_new_2015}
K.~Javed, R.~Gouriveau, N.~Zerhouni, A new multivariate approach for
  prognostics based on extreme learning machine and fuzzy clustering, {IEEE}
  Transactions on Cybernetics 45~(12) (2015) 2626--2639.

\bibitem{eker2014similarity}
{\"O}.~F. Eker, F.~Camci, I.~K. Jennions, A similarity-based prognostics
  approach for remaining useful life prediction, in: Prognostics and Health
  Management Society. 2nd European Conference of the, PHM Society, 2014, pp.
  1--5.

\bibitem{ramasso_performance_2014}
E.~Ramasso, A.~Saxena, Performance benchmarking and analysis of prognostic
  methods for {CMAPSS} datasets., International Journal of Prognostics and
  Health Management 5~(2) (2014) 1--15.

\bibitem{babu_deep_2016}
G.~S. Babu, P.~Zhao, X.-L. Li, Deep convolutional neural network based
  regression approach for estimation of remaining useful life, in:
  International conference on database systems for advanced applications,
  Springer, 2016, pp. 214--228.

\bibitem{heimes_recurrent_2008}
F.~O. Heimes, Recurrent neural networks for remaining useful life estimation,
  in: Prognostics and Health Management, 2008. {PHM} 2008. International
  Conference on, {IEEE}, 2008, pp. 1--6.

\bibitem{zheng_long_2017}
S.~Zheng, K.~Ristovski, A.~Farahat, C.~Gupta, Long short-term memory network
  for remaining useful life estimation, in: Prognostics and Health Management
  ({ICPHM}), 2017 {IEEE} International Conference on, {IEEE}, 2017, pp. 88--95.

\bibitem{malhotra_multi-sensor_2016}
P.~Malhotra, V.~{TV}, A.~Ramakrishnan, G.~Anand, L.~Vig, P.~Agarwal, G.~Shroff,
  Multi-sensor prognostics using an unsupervised health index based on lstm
  encoder-decoder, {arXiv} preprint {arXiv}:1608.06154.

\bibitem{wang_trajectory_2010}
T.~Wang, Trajectory similarity based prediction for remaining useful life
  estimation, Ph.D. thesis, University of Cincinnati (2010).

\bibitem{lin_riemannian_2008}
T.~Lin, H.~Zha, Riemannian manifold learning, {IEEE} Transactions on Pattern
  Analysis and Machine Intelligence 30~(5) (2008) 796--809.

\bibitem{khan_universal_2016}
F.~Khan, D.~Kari, I.~A. Karatepe, S.~S. Kozat, Universal nonlinear regression
  on high dimensional data using adaptive hierarchical trees, {IEEE}
  Transactions on Big Data 2~(2) (2016) 175--188.

\bibitem{xie_online_2013}
Y.~Xie, R.~Willett, Online logistic regression on manifolds, in: 2013 IEEE
  International Conference on Acoustics, Speech and Signal Processing, 2013,
  pp. 3367--3371.

\bibitem{xie_change-point_2013}
Y.~Xie, J.~Huang, R.~Willett, Change-point detection for high-dimensional time
  series with missing data, {IEEE} Journal of Selected Topics in Signal
  Processing 7~(1) (2013) 12--27.

\bibitem{wang_adaptive_2005}
Z.~Zhang, J.~Wang, H.~Zha, Adaptive manifold learning, IEEE Transactions on
  Pattern Analysis \& Machine Intelligence 34~(2) (2012) 253--265.

\bibitem{liu_users_2012}
Y.~Liu, D.~K. Frederick, J.~A. {DeCastro}, J.~S. Litt, W.~W. Chan, User's guide
  for the commercial modular aero-propulsion system simulation (c-{MAPSS}):
  Version 2, NASA/TM-pp.2012–217432[R].

\end{thebibliography}

\end{document}